\documentclass[a4paper,11pt]{article}
\usepackage{jcappub} 
\usepackage{lineno}

\usepackage[normalem]{ulem}
\usepackage{xcolor}

\title{Phenomenology of DSR-relativistic in-vacuo dispersion in FLRW  spacetime}

\author[a,b]{G. Amelino-Camelia}
\author[a,b]{D. Frattulillo}
\author[a,b]{G. Gubitosi}
\author[c]{G. Rosati}
\author[d]{S. Bedić}
\affiliation[a]{Dipartimento di Fisica Ettore Pancini, Universit\`a di Napoli ``Federico II''}
\affiliation[b]{INFN, Sezione di Napoli, 
Complesso Univ. Monte S. Angelo, I-80126 Napoli, Italy}
\affiliation[c]{Institute for Theoretical Physics, University of Wroc{\l}aw, Pl. Maksa Borna 9, Pl–50-204 Wroc{\l}aw, Poland}
\affiliation[d]{ICRANet
, P.le della Repubblica 10, 65100 Pescara, Italy and\\
ICRA and University of Rome ``Sapienza'', Physics Department, P.le
A. Moro 5, 00185 Rome, Italy}

\abstract{Studies of in-vacuo dispersion are the most active area of quantum-gravity phenomenology.
The way in which in-vacuo dispersion produces redshift-dependent  corrections to the time of flight of astrophysics particles depends on the model-dependent interplay between Planck-scale effects and spacetime curvature/expansion, and we here derive  the most general formula for the leading order redshift-dependent correction to the time of flight  for the scenario in which relativistic symmetries are deformed at the Planck scale (DSR) for the constant-curvature case.
We find that, contrary to the broken symmetries scenario (LIV), where in principle any arbitrary form of redshift dependence could be allowed, for the DSR scenario only linear combinations of three possible forms of redshift dependence are allowed.
We also derive a generalization of our results to the FRW case, and 
discuss some specific combinations of the three forms of redshift dependence 
whose investigation might deserve priority from the quantum-gravity perspective. 
}

\makeatletter
\gdef\@fpheader{}
\makeatother

\begin{document}
\maketitle
\flushbottom

\newpage

\section{Introduction}
The possibility of Planck-scale departures from (local) Lorentz invariance arises in several quantum-gravity proposals (see, {\it e.g.}, Refs.\cite{GACreview, MattinglyReview,COSTReview} and references therein).
In some scenarios (LIV) relativistic invariance  
 is broken~\cite{GRBnature,Alfaro:1999wd,Gambini:1998it}, giving rise to a preferred frame, while in other scenarios (DSR)
 relativistic invariance is merely deformed~\cite{Amelino-Camelia:2000stu,Magueijo:2002am,Kowalski-Glikman:2002iba}, preserving the equivalence of reference frames but requiring a deformation of relativistic laws of transformation
among observers.

We are here concerned with tests of the fate of relativistic symmetries in quantum gravity which are based on 
time-of-flight measurements, and when analyzed in a flat spacetime the implications of the broken-symmetry and the deformed-symmetry scenario are indistinguishable: in both
cases one gets a leading correction $\Delta t$ to the special-relativistic time of flight which (assuming linear dependence on the energy $E$ of the particle) is
governed by~\cite{GRBnature}
\begin{equation}
    \Delta t= \eta \frac{E}{M_{Pl}} T\,,
\end{equation}
where $T$ is the (time) distance of the source, $\eta$ is a phenomenological parameter and $M_{Pl}$ is the Planck scale.
However, matters become more complicated 
(and the differences between the
broken-symmetry and the deformed-symmetry scenario become more tangible) if
one takes into account the expansion of spacetime: the interplay between quantum-gravity effects and curvature of spacetime can produce several alternative forms of redshift dependence of the effect.
In the LIV broken-symmetry scenario one has no constraints from symmetries and in principle any arbitrary form of redshift dependence could be allowed (see for instance Refs.~\cite{PiranMartinez,Pfeifer:2018pty});  however, LIV-based data analyses all rely on a particular form of redshift dependence introduced
in Ref.~\cite{JacobPiran} 
(also see Ref.~\cite{EllisMavroFRWdelay})
which gives the following redshift dependence
\begin{equation}
\label{JacobPiran}
    \Delta t= \eta \frac{ E}{M_{Pl}}\int_0^z \frac{1+\bar z}{H(\bar z)}d\overline{z}\, ,
\end{equation}
where $z$ is the redshift of the source, related to the scale factor $a(t)$ as $z(t) = 1/a(t)-1$, $H(z)$ is the Hubble parameter, that for the $\Lambda$CDM model is~\footnote{$\Omega_\Lambda$, $H_0$ and $\Omega_m$ denote,
respectively, the cosmological constant, the Hubble constant, and the matter fraction, for which we take the values given in Ref.~\cite{Planck:2018vyg}.} $H(z)=H_0\sqrt{\Omega_\Lambda + (1+z)^3 \Omega_m}$. \newline
In the DSR deformed-symmetry scenario 
the possibilities for the interplay between quantum-gravity effects and curvature of spacetime are significantly limited by the requirement that the merging picture should be compatible with (however deformed) relativistic invariance. In a previous study~\cite{DSRFRW}
(also see Ref. \cite{Bolmont:2022yad})
two examples of DSR-compatible forms of redshift dependence were identified.

In the study we are here reporting we establish what is the most general form of redshift dependence allowed by the requirement of DSR compatibility.
We find that, in addition to the two forms of redshift dependence already previously identified~\cite{DSRFRW}, there is only a third possible form of redshift dependence. Of course, also linear combinations of these three possible forms of redshift dependence are allowed.

Our analysis starts (Sec.~\ref{sec:deformation}) by considering the simple case of propagation in a de Sitter spacetime, where the possible DSR-relativistic scenarios can be characterized fully in terms of deformations of the symmetries of de Sitter spacetime, which are described by an algebra of 10 generators (spacetime translations, rotations and boosts)~\cite{Marciano:2010gq,DSRdS}. We establish that there are only three different terms that can give a DSR-compatible description of the redshift dependence of the time of flight. These results are generalized in Sec.~\ref{sec:timedelay} to the case of particles propagating in a FLRW spacetime through a slicing procedure, whose robustness was well tested already in Refs. \cite{Marciano:2010gq,DSRFRW}.
In Section~\ref{sec:specialCases} we describe the phenomenology of some specific combinations of the three DSR-compatible forms of redshift dependence which might be particularly significant from the quantum-gravity perspective.
In the closing Section~\ref{conclusions} we offer a perspective on our results and on possible further developments for this research programme.

We use natural units $c=\hbar=1$.

\section{Most general deformation of the de Sitter algebra of symmetries}
\label{sec:deformation}

As announced, our analysis takes off from an investigation of the most general Planck-scale deformation 
of the de Sitter algebra. We denote by $H$ the curvature parameter of the de Sitter algebra and we denote by $\ell$
(a length scale assumed to be of the order of the Planck length)
the deformation parameter and we shall be satisfied working at leading order 
in $\ell$.

{Working in 1+1 spacetime dimensions, we start by characterizing the most general deformation of the mass Casimir.
We ask that the limits for vanishing curvature ($H\rightarrow 0$) and vanishing deformation ($\ell \rightarrow 0$) are well-defined and in particular that the latter leaves us with the standard de Sitter Casimir.  Moreover,   we require that the vectorial properties of the generators are accounted for, so that the generalization to higher spatial dimensions does not affect space-rotational invariance.
The most general deformation of the de Sitter Casimir which satisfies these requirements is:
\begin{equation}
{\cal C}=E^{2}-p^{2}-2HNp+\ell\left(\alpha E^{3}+\beta Ep^{2}+2\gamma HNEp+4\mu H^{2}N^{2}E\right) .
\label{casimir}
\end{equation}
Here $\alpha,\beta,\gamma,\mu$ are dimensionless parameters.
Compared to previous studies~\cite{DSRdS,DSRFRW} of Planck-scale deformations of the de Sitter Casimir,
our Eq.~(\ref{casimir})
includes  two additional terms,
the one parametrized  by $\gamma$ (which however had been considered in Ref.~\cite{Barcaroli:2015eqe} in the context of a study of particle kinematics with $q$-de Sitter Hopf-algebra symmetries) and
the one parametrized by $\mu$. 
While the Casimir \eqref{casimir} is general and does not come from a specific quantum gravity model, one can interpret the terms proportional to $H\ell$ 
 in the framework of a ``quantum group" $q$-deformation of the Poincaré algebra, where the deformation is triggered by a combination of the curvature scale and the ``quantum gravity" scale encoded in the parameter $q$ (see, e.g., Refs.~\cite{GACSmolinqdeSitter,Barcaroli:2015eqe}). 
These theory implications may well deserve dedicated studies, but we here intend to focus on the issues relevant for phenomenology.

The most general algebra of symmetry generators/charges that leaves the Casimir~(\ref{casimir}) invariant can be described by the following set of Poisson brackets (see also Refs.~\cite{DSRdS,DSRFRW}):
\begin{equation}
\begin{gathered}\left\{ E,p\right\} =Hp-\ell HE\left[\left(\alpha+\gamma-\sigma\right)p+4\mu HN\right],\\
\left\{ N,E\right\} =p+HN-\ell E\left[\left(\alpha+\beta-\sigma\right)p+HN\left(\alpha+\gamma-\sigma\right)\right],\\
\left\{ N,p\right\} =E+\frac{\ell}{2}\left[\left(\alpha+2\sigma\right)E^{2}+\beta p^{2}+2\gamma HNp+4\mu H^{2}N^{2}\right] .
\end{gathered}
\label{algebra}
\end{equation}
These define a deformation of the standard de Sitter algebra.
Notice that, in addition to the parameters characterizing deformations of the Casimir, the algebra admits the additional numerical parameter $\sigma$.

So, at this kinematical level,  departures from the standard relativistic symmetries are characterized by five independent parameters. However, as we shall here show, the implications for time-of-flight measurements only involve three independent combinations of these 5 parameters.

\section{In-vacuo dispersion for DSR-FLRW scenarios }
\label{sec:timedelay}

In this section we use the description of deformed de Sitter symmetries given in the previous section in order to derive the time-delay formulas which are here of interest. Those formulas are then generalized to the FLRW case using, as announced, the ``slicing" technique.

\subsection{Time delays for deformed de Sitter spacetimes}\label{sec:DSRdS}

In deriving our time-delay results, we must keep safe from possible relativistic artifacts due to the relativity of locality~\cite{taming,principle,kbob,Mignemi:2019yzn} which is present in this deformed-relativistic scenarios. We accomplish that by relying on two observers, one local to the emission event and the other one local to the detection event.
Following the  strategy of analysis outlined in~\cite{DSRdS,DSRFRW, Barcaroli:2015eqe, Amelino-Camelia:2013uya}, we perform a finite translation that allows us to express the coordinates of an observer at the detector $(t^B,x^B)$ in terms of the coordinates of an observer at the source $(t^A,x^A)$, defined by the prescription
\begin{equation}
(t^B,x^B)=e^{-\xi p}\triangleright e^{-\zeta E}\triangleright (t^A,x^A),
\end{equation}
where $\triangleright$ stands for the action by Poisson bracket of the corresponding generators\footnote{For a generator $G$ with parameter $a$, the finite action on a coordinate $x$ is $e^{{a}G} \triangleright x \equiv \sum_{n=0}^{\infty} \frac{a^n}{n!}\left\lbrace G,x\right\rbrace_n $, where $ \lbrace G,x \rbrace_n = \lbrace G, \lbrace G,x  \rbrace_{n-1}\rbrace$, $\lbrace G, x \rbrace_0 = x $.
In this formalism, the composed action of a spatial translation followed by a time translation is given by $e^{-\xi p}\triangleright e^{-{\zeta}E}\triangleright x$.} and $\xi$ and $\zeta$ are respectively the space and time translation parameters.
We then find that  two photons with energy difference $\Delta E$ at the detector, emitted simultaneously by a distant source, reach the detector with a time difference 
\begin{equation}\label{dstimedelayH}
\Delta t=\ell \Delta E\left(\left(\beta-\gamma+\sigma+\mu\right)\frac{e^{2HT}-1}{2H}+\left(\alpha+\gamma-\sigma-2\mu\right)T+\mu\frac{1-e^{-2HT}}{2H}\right),
\end{equation}
where $T$ is the comoving (time)  distance between the source and the detector.
In terms of the redshift of the source $z=e^{HT}-1$ this reads
\begin{equation}\label{dstimedelay}
\Delta t  = \! \frac{\ell \Delta E}{H} \!\left( \!\! \left(\beta \!-\! \gamma \!+\! \sigma \!+\! \mu\right) \!\!\left( \! z \!+ \! \frac{z^{2}}{2}\right) + \left(\alpha \!+\! \gamma \!-\! \sigma \!-\! 2\mu\right)\ln \! \left(1 \!+\! z\right)+\mu \! \left( \!\frac{z+z^{2}/2}{1+2z+z^{2}}\! \right) \!\! \right) \!.
\end{equation}
As announced, we are finding that the five numerical parameters that characterize the deformation of the kinematics in Eqs.~\eqref{casimir}-\eqref{algebra} combine to produce only three different terms characterizing the functional dependence of the time delay on the redshift. 
In particular, of the two terms in~(\ref{casimir}) that were not considered in~\cite{DSRFRW,DSRdS}, the one parameterized by $\gamma$ does not add to the time delay formula any new functional dependence on the redshift compared to what was already considered in~\cite{DSRFRW,DSRdS}. On the other hand, the new term in 
the dispersion relation parameterized by $\mu$ produces a functional dependence on the redshift that was not considered before.

\subsection{DSR-FLRW time delays}
\label{sec:DSRFLRWtimedelay}
Up to this point we rigorously showed that in-vacuo dispersion in (DSR-)relativistic quantum de Sitter spacetimes can only be characterized by (linear combinations of) certain 3 independent forms of redshift dependence. This analysis did not require us to make additional assumptions besides the quantum de Sitter invariance.
However, for what concerns the phenomenology of in-vacuo dispersion, we cannot rely on the constant-curvature assumption that applies to de Sitter  spacetime:
we need
  to generalize our results to an FLRW expanding spacetime, but a (DSR-)relativistic FLRW quantum geometry has still not been developed}.
The ultimate goal would be to have some generalization of Einstein's equation applicable to
quantum geometries, but that looks still like a distant goal for quantum-spacetime research.

We shall rely on a semiheuristic approach, which makes the reasonable assumption that the relationship between travel times in quantum de Sitter and FLRW spacetimes should preserve
at least some aspects of the
structure of the corresponding relationship between travel times in 
classical de Sitter and FLRW spacetimes. Our strategy of analysis can be better appreciated
by taking as reference the successful semiheuristic approach of Refs. \cite{JacobPiran,PiranMartinez}
which led to the identification of
the Jacob-Piran redshift dependence, which is the standard of reference 
for LIV (broken-relativistic-symmetry) quantum-spacetime phenomenology.
Starting from the results on in-vacuo dispersion in LIV flat quantum spacetimes, 
 one could contemplate any arbitrary form of redshift dependence in a LIV FLRW quantum spacetime, since the LIV
case provides no  relativistic symmetry constraints. Indeed in Ref. \cite{PiranMartinez}
some alternative forms of redshift dependence were considered. The Jacob-Piran redshift dependence was singled out through the assumption~\cite{JacobPiran,PiranMartinez} that, 
in LIV quantum spacetimes, momenta should be affected by redshift in the same way that they do in classical general-relativistic spacetime. It is turning out that this assumption is correct only in a subset of quantum-spacetime models: the interplay between spacetime expansion and quantum properties of spacetime often produces a modification of the effect of redshift on
momenta (see, {e.g.}, \cite{Pfeifer:2018pty,Lobo:2016xzq}).

To see how our strategy of analysis is related to the Jacob-Piran approach we observe that
in the ordinary general-relativistic case travel times in FLRW can be obtained from travel
times in de Sitter equivalently assuming that momenta redshift general-relativistically and
assuming that the travel time in FLRW is obtained by ``de Sitter slicing", {\it i.e.}
describing propagation in FLRW, with its time-dependent   expansion rate $H(t)=\dot{a}(t)/a(t)$, as a sequence of infinitesimal steps of propagation in de Sitter spacetime with scale factor $a(t)=\exp(Ht)$
(see Appendix \ref{appendix}). Our assessment is that in 
the relativistic quantum spacetimes whose phenomenology we are describing
the status of ``de Sitter slicing" is much safer than that of the effect of redshift of
momenta, and we shall therefore rely on de Sitter slicing. 
For this we follow exactly the procedure described in detail in Ref.~\cite{DSRFRW} (also see Appendix \ref{appendix}), in which the propagation of signals in a  FLRW spacetime with deformed local relativistic symmetries is described by defining a  sequence of intermediate observers along the particle's trajectory, such that each observer is local to the particle at a given spacetime point. Propagation of the signal between two such nearby observers is described by using the  deformed de Sitter kinematics as done in the previous subsection and contributes to the total travel time  by an amount given by~(\ref{dstimedelayH}). The full trajectory (and the corresponding total travel time) in the deformed FLRW spacetime is  reconstructed by suitably matching~\cite{DSRFRW} the observations made by subsequent observers and considering a limiting procedure in which the  number of intermediate observers is sent to infinity, while decreasing their distance to zero.

Following this procedure we find that from our Eq.(\ref{dstimedelay}) the time delay in the deformed FLRW case reads:
\begin{equation}\label{frwdelayA}
\begin{split}
\Delta t =  \frac{\Delta E}{M_{Pl}} \int_{0}^{z}\frac{d\overline{z}\left(1 \!+\! \overline{z}\right)}{H\left(\overline{z}\right)}\Bigg[& \eta_1 +  \eta_2\left( \! 1 \!-\! \left( \! 1 \!-\! \frac{H\left(\overline{z}\right)}{1 \!+\! \bar{z}}\int_{0}^{\overline{z}} \! \frac{d\overline{z}'}{H\left(\overline{z}'\right)}\right)^{\!2}\right) \\ & 
+ \eta_3\left(\! 1 \!-\! \left( \!1 \!-\! \frac{H\left(\overline{z}\right)}{1 \!+\! \bar{z}}\int_{0}^{\overline{z}}\! \frac{d\overline{z}'}{H\left(\overline{z}'\right)}\right)^{\!4}\right) \Bigg] .
\end{split}
\end{equation}

As we anticipated, the time delay depends on only three numerical parameters $\eta_1$, $\eta_2$, and $\eta_3$ that are  to be determined by experiments.
Their relation with the parameters of the deformed algebra introduced in Sec.~\ref{sec:deformation} is given by 
$$\eta_1= (\alpha+\beta) \,,\,\,\,\,\,\, \eta_2= (-\alpha-\gamma+\sigma+2\mu) \,, \,\,\,\,\,\,\, \eta_3=-\mu \, .$$
Of course, for the case in which the
$H(z)$ is actually redshift independent the FLRW picture turns into a de Sitter picture and our result 
(\ref{frwdelayA}) reproduces our result (\ref{dstimedelay}).

\begin{figure}[h!]
\centering
\includegraphics[width=0.7 \textwidth]{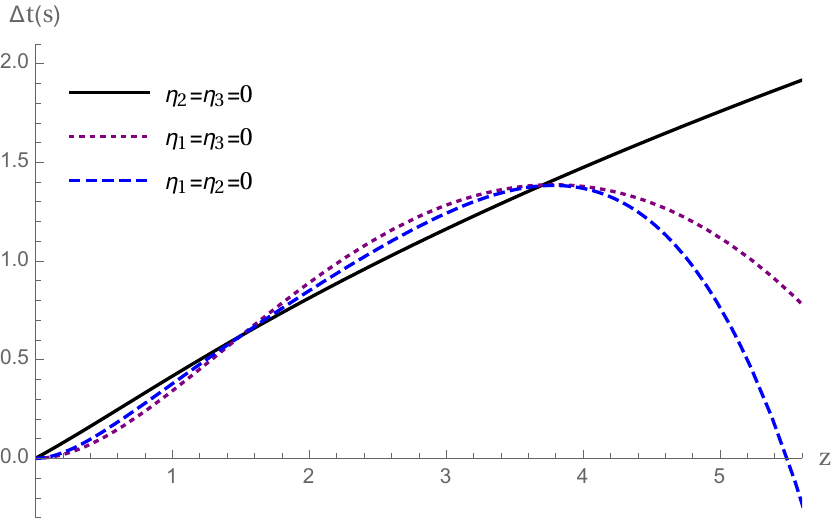}
\caption{{The redshift dependence of the three terms contributing to the time delay in Eq.~(\ref{frwdelayA}). Each curve corresponds to the time delay~(\ref{frwdelayA}) when only one of the three parameters $\eta_1$, $\eta_2$ and $\eta_3$ is different from zero.
The continuous black line assumes that $\eta_1=1$ (while $\eta_2=\eta_3=0$)  and $\Delta E = 10~\text{GeV}$.
The dotted purple and dashed blue lines still assume $\Delta E = 10~\text{GeV}$, and they are obtained by fixing, respectively,  $\eta_2$ and $\eta_3$ so that the time delay matches the one of the black continuous curve at $z=1.5$.
}
}
\label{figura3param}
\end{figure}

{In Figure~\ref{figura3param} we plot the redshift dependence  of the three terms in~(\ref{frwdelayA}).}

While any linear combination of the three redshift-dependent terms in~(\ref{frwdelayA}) is a good candidate for a DSR-FLRW time-delay formula,
we find that the parametrization in terms of $\eta_1$, $\eta_2$, and $\eta_3$ turns out to be convenient for the comparison of some specific phenomenological scenarios { that we are going to discuss in the next section}.
In particular, when $\eta_2=\eta_3=0$, we are left with the term parametrized by $\eta_1$ which gives the same time delay that was obtained in the LIV scenario by Jacob and Piran  in Ref.~\cite{JacobPiran}.
On the other hand, scenarios with vanishing  $\eta_1$ or $\eta_2$ characterize, respectively,  two noteworthy cases that we shall discuss in the following:
when $\eta_1$ vanishes one obtains curvature-induced scenarios (see Subsec.~\ref{sec:curvInduced}), while vanishing $\eta_2$ relates to theoretical models where energies add up trivially (see Subsec.~\ref{sec:trivialE}).


\section{Some noteworthy special cases of DSR-FLRW time delay}
\label{sec:specialCases}

We have shown that the requirement of compatibility with DSR-relativistic invariance limits to combinations of only 3 independent forms of redshift dependence 
for time-delay phenomenology. 
Still, even just a 3-parameter formula is a rather wide ``hunting field" for the phenomenology of time delays in astrophysics, where data are scarce and often of poor quality. In this section we attempt to motivate from the theoretical perspective some specific choices of the parameters $\eta_1$, $\eta_2$, and $\eta_3$ in~(\ref{frwdelayA}) which might deserve being the ``first targets" for the phenomenology.

\subsection{Curvature-induced scenarios}
\label{sec:curvInduced}

A  scenario that so far received little attention in the literature, but that could have interesting phenomenological implications, is the one where the quantum gravity effects are triggered by spacetime curvature. This is a scenario where the interplay between curvature effects and Planck-scale effects produces results that are the most distant from what one would guess based on analysing Planck-scale effects in the flat-spacetime approximation.  Indeed, in this scenario in-vacuo dispersion occurs only in combination with spacetime curvature/expansion so that when curvature is negligible there is no expected time delay.
Theoretically, these scenarios find motivation in some studies based on a Hopf-algebra description of the symmetries of quantum spacetime~\cite{GACSmolinqdeSitter,aschieri09}, as well as in some considerations arising from loop-quantum-gravity research~\cite{bianchiRovelli}.
A first study of these ``curvature-induced" scenarios was performed in~\cite{curvInducedLIV}, relying on a toy model where relativistic symmetries are broken.
The preliminary results reported in~\cite{curvInducedLIV}, confronting the slow onset of the quantum gravity effects in the FLRW time-delays, that are typically expected in curvature-induced scenarios, with data relative to gamma-ray-burst observations, showed how these features might have interesting implications for experimental studies. 

Here we show that there is a choice of parameters that produces a curvature-induced time-delay also in the DSR-FLRW framework we constructed in the previous sections.
As explained in~\cite{curvInducedLIV}, the requirement for having only curvature-induced terms in the time-delay formula amounts to asking that the coefficient of the first-order term in an expansion around $z=0$ of the expression~(\ref{frwdelayA}) vanishes.
Indeed, expanding the redshift formula $z(t)=1/a(t)-1$ for small distances (i.e. small (negative) times $t=-T$), one gets $z(-T)\simeq  H_0 T$, where the Hubble constant is defined as $H_0=\frac{1}{a}\frac{da}{dt}|_{t=0}$. It follows that terms linear in $z$ in the expansion of $\Delta t$ will be proportional to $\frac{\Delta E}{M_{Pl}}\frac{z}{H_0} \simeq \frac{\Delta E}{M_{Pl}} T $, and will survive even in the absence of spacetime curvature. Thus, only terms involving powers of $z$ higher than 1 contribute to curvature-induced time-delay effects.

The leading order expansion in terms of the redshift of Eq.~(\ref{frwdelayA}) gives
\begin{equation}
   \Delta t\simeq \frac{\Delta E}{M_{Pl}H_0}\left( \eta_1 z + O(z^2)\right).
\end{equation}
Setting to zero the first order term corresponds to imposing the constraint $\eta_1=0$ (i.e. $\alpha=-\beta$ in terms of the kinematical parameters).
Notice also that the same condition is obtained in the DSR-de Sitter case of Subsec.~\ref{sec:DSRdS} if one asks that the time delay of Eqs.~\eqref{dstimedelayH} and~\eqref{dstimedelay} vanishes in the limit of vanishing spacetime curvature $H$. Indeed, considering the limit $H\to 0$ in \eqref{dstimedelayH} (or in~(\ref{dstimedelay}), noticing that $z\simeq H T + O(H^2 T^2)$, we obtain
\begin{equation}
    \Delta t = \ell \frac{\Delta E}{H} \left((\alpha+\beta)z + O(z^2)\right) = \ell \Delta E \left((\alpha+\beta)T + O(H T^2)\right),
\end{equation}
which  gives again the condition $\alpha=-\beta$, i.e. $\eta_1=0$.

By imposing the condition $\eta_1=0$ in \eqref{frwdelayA} the time delay expression reduces to 
\begin{equation}
\label{curvinddelay}
\begin{split}
    \Delta t =  \frac{\Delta E}{M_{Pl}} \int_{0}^{z}\frac{d\overline{z}\left(1+\overline{z}\right)}{H\left(\overline{z}\right)}
    \Bigg[ &~ \eta_2\left(1 \!-\! \left(1 \!-\! \frac{H\left(\overline{z}\right)}{1+\bar{z}}\int_{0}^{\overline{z}}\frac{d\overline{z}'}{H\left(\overline{z}'\right)}\right)^{2}\right) \\& 
    + \eta_3\left(1-\left(1-\frac{H\left(\overline{z}\right)}{1+\bar{z}}\int_{0}^{\overline{z}}\frac{d\overline{z}'}{H\left(\overline{z}'\right)}\right)^{4}\right)\Bigg] .
    \end{split}
\end{equation}
This formula, in which only two independent parameters appear, describes the most general curvature-induced in-vacuo dispersion scenario arising from the deformation of symmetries under the hypotheses of Secs.~\ref{sec:deformation} and~\ref{sec:timedelay}.

\subsection{Scenarios with undeformed addition of energy}
\label{sec:trivialE}

Apart from the deformation of the mass Casimir and the algebra of relativistic symmetry generators, another important ingredient of DSR models concerns the conservation law of energy-momenta for processes involving multiple particles.
In order for the conservation law to be invariant under the deformed symmetries, it  must be accordingly deformed~\cite{Amelino-Camelia:2000stu,Amelino-Camelia:2011gae}.

In order to study the possible deformations of the energy-momenta conservation law for the deformed de Sitter scenario described in Sec.~\ref{sec:deformation}, we consider the total energy and momentum charges resulting from the composition law in the two-particle case.
Keeping the description as general as possible, we consider all the possible terms that can be added to the standard (linear) special relativistic sum law of energy-momenta at leading order in the deformation parameter $\ell$. This is found by requiring that the total charges close the same algebra~\eqref{algebra} as the single particle energy and momenta (this ensures the relativistic properties of the composition law) and that no deformation terms which involve only one particle charge are present, so that we recover the definition of single particle charge when the charges of the second particle are zero~\cite{Amelino-Camelia:2013sba,Amelino-Camelia:2023rkg}. Moreover, we require the same conditions of analyticity, dimensional consistency, and ``vectorial properties" adopted for the algebra deformation in Sec.~\ref{sec:deformation}.
The most general composition laws complying with these requirements is given by the following:
\begin{equation}\label{complaw1}
    \begin{aligned}
    &E_{tot}=E_1+E_2+\ell ((2\sigma-\beta-a-b) P_1P_2+(c-\gamma+\sigma) H (N_1P_2+P_1N_2) \\& ~~~~~~~~~ -\alpha E_1E_2+2(c-2\mu)H^2N_1N_2 )\\
    &P_{tot}=P_1+P_2+\ell\left((\sigma-b) E_1P_2+(\sigma-a)  E_2P_1+cH(N_1E_2+E_1N_2) \right)\\
&N_{tot}=N_1+N_2+\ell\left(aE_1N_2+bE_2N_1\right)\,.
    \end{aligned}
\end{equation}
Notice that three additional parameters $(a,b,c)$, that didn't appear in~(\ref{algebra}), are allowed.

While all possibilities contemplated by our Eq.~(\ref{complaw1}) deserve being investigated, we feel that priority should be given to scenarios in which 
the addition law of particle energies remains undeformed.
This is suggested by experience \cite{Amelino-Camelia:2013sba,Amelino-Camelia:2011gae}
with the implications of these modified addition laws
in which one finds that preserving the linearity of addition of energies is advantageous from the point of view of the interpretation of the results.
Moreover this requirement finds further motivation in scenarios where the DSR framework can be associated with a quantum group deformation of de Sitter symmetries, where the summation law of the charges/generators corresponds to a ``coproduct rule" of the Hopf-algebra generators. In that case an undeformed summation law of energies would correspond to a ``primitive coproduct" for energy/time-translation generators, that is necessary for having a ``time-like" q-deformation of de Sitter symmetries~\cite{Ball19943D,Ballesteros:2014kaa,Ballesteros:2017pdw,jack3Dgravity}.

The requirement for the composition of energy to be undeformed imposes the following constraints between the kinematical parameters:
\begin{equation}
    \alpha=0\qquad \gamma-\sigma=2\mu \ ,
\end{equation}
amounting to $\eta_2=0$ in~\eqref{frwdelayA}.
The expression for time delay in the deformed FLRW scenario then becomes
\begin{equation}
        \Delta t=\frac{\Delta E}{M_{Pl}} \int_{0}^{z}\frac{d\overline{z}\left(1+\overline{z}\right)}{H\left(\overline{z}\right)}\left[\eta_1+\eta_3\left(1-\left(1-\frac{H\left(\overline{z}\right)}{1+\bar{z}}\int_{0}^{\overline{z}}\frac{d\overline{z}'}{H\left(\overline{z}'\right)}\right)^{4}\right)\right]\,,
\end{equation}
in which, again, only two independent parameters appear.

\subsection{A one-parameter scenario: curvature-induced and undeformed addition of energy}
\label{sec:oneparam}

Combining the requirements of Subsecs.~\ref{sec:curvInduced} and~\ref{sec:trivialE} we obtain a scenario that has only one free numerical parameter to be determined by experiments, $\eta_3$, and is characterized by undeformed composition law of energies and a curvature-induced time delay effect. 

\begin{figure}[h!]
\centering
\includegraphics[width=0.7 \textwidth]{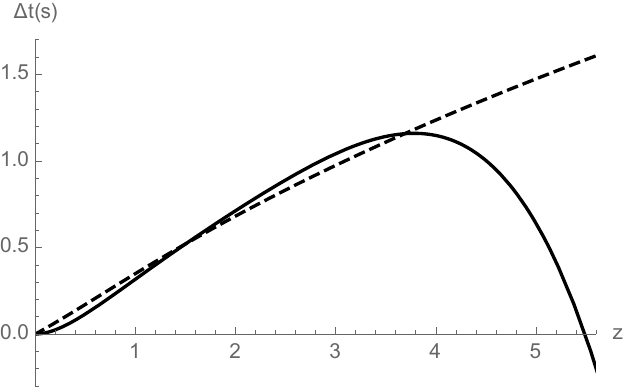}
\caption{The continuous line corresponds to the one-parameter scenario described by~(\ref{frwdelayOneparam}), with $\Delta E = 10~\text{GeV}$ and $\eta_3=1$.
The dashed line represents the expected time delay for the Jacob-Piran case~(\ref{JacobPiran}), corresponding to setting $\eta_2=\eta_3=0$ in~(\ref{frwdelayA});
the remaining free parameter in~(\ref{frwdelayA}), $\eta_1$, is fixed by asking that  the two lines  cross at $z=1.5$. Also for the dashed line we use $\Delta E = 10~\text{GeV}$.}
\label{figura1}
\end{figure}

The resulting formula for the time delay is 
\begin{equation}
\label{frwdelayOneparam}
    \Delta t= \eta_3 \frac{\Delta E}{M_{Pl}} \int_{0}^{z}\frac{d\overline{z}\left(1+\overline{z}\right)}{H\left(\overline{z}\right)}\left[1-\left(1-\frac{H\left(\overline{z}\right)}{1+\bar{z}}\int_{0}^{\overline{z}}\frac{d\overline{z}'}{H\left(\overline{z}'\right)}\right)^{4}\right] .
\end{equation}
In Figure~\ref{figura1} we compare the redshift dependence described by this formula to the one of the Jacob-Piran scenario~(\ref{JacobPiran}).

\subsection{Alternative picture with the time delay changing sign}
\label{sec:signchange}

It is rather noteworthy that in our one-parameter scenario which is curvature induced and is compatible with undeformed addition of energy
the time delay changes sign at high redshift.
So far all the scenarios motivated in the literature gave rise to monotonic dependence of the time delay on redshift, and it is interesting that our one-parameter scenario, with its appealing theoretical qualities, is not monotonic. This led us also to investigate how frequently in our 3-dimensional parameter space such changes of the sign of the time delay occur and what sort of functional dependence on redshift are then found in such cases. We found that cases in which the time delay changes sign are not at all exceptional, and a variety of forms of dependence on redshift can be found. 

\begin{figure}[h!]
    \centering
\includegraphics[width=0.7 \textwidth]{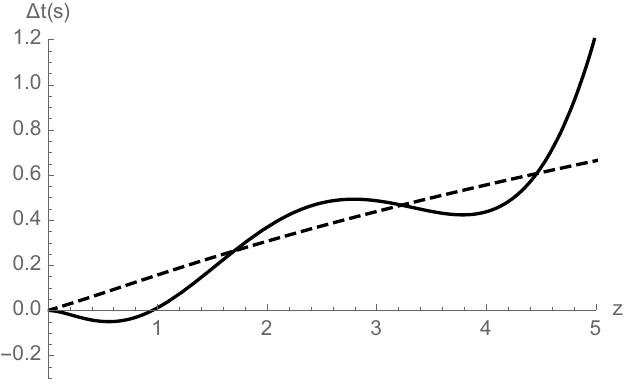}
    \caption{The continuous line corresponds to the ``curvature-induced" scenario described by~(\ref{curvinddelay}) with $\eta_2=4$ and $\eta_3=-3$, see Eq.~(\ref{changesign}).
The dashed line represents the expected time delay for the Jacob-Piran case~(\ref{JacobPiran}), corresponding to setting $\eta_2=\eta_3=0$ in~(\ref{frwdelayA}); the remaining free parameter in~(\ref{frwdelayA}),
$\eta_1$, is fixed by asking that  the two lines  cross at $z=1.7$ (which allows for a particularly interesting comparison). As for Figure~\ref{figura1}, we set $\Delta E = 10~\text{GeV}$ for both lines.} \label{fig:fig2}
\end{figure}

As an illustrative example we focused on the case of effects which are curvature induced ($\eta_1=0$) and with $\eta_2=4,\eta_3=-3$. With this choice, Eq.~(\ref{frwdelayA}) becomes
\begin{equation}\label{changesign}
\begin{split}
\Delta t = \frac{\Delta E}{M_{Pl}} \int_{0}^{z}\frac{d\overline{z}\left(1+\overline{z}\right)}{H\left(\overline{z}\right)} \Bigg[ & 4\left(1 \!-\! \left(1 \!-\! \frac{H\left(\overline{z}\right)}{1+\bar{z}}\int_{0}^{\overline{z}}\frac{d\overline{z}'}{H\left(\overline{z}'\right)}\right)^{2}\right) \\& -3\left(1 \!-\! \left(1 \!-\! \frac{H\left(\overline{z}\right)}{1+\bar{z}}\int_{0}^{\overline{z}}\frac{d\overline{z}'}{H\left(\overline{z}'\right)}\right)^{4}\right) \Bigg] .
\end{split}
\end{equation}
 As shown in Figure~\ref{fig:fig2},
 in this scenario the redshift dependence starts off at small redshifts with opposite sign with respect to the Jacob-Piran scenario, but then for redshift greater than 1 
 (up to redshift of about 4.5) approximates reasonably well (oscillating around it) the Jacob-Piran scenario. It would therefore be a valuable aspect of maturity of this phenomenology when the quality of data at high redshift will prove to be sufficient for discriminating between this scenario and the Jacob-Piran scenario.
 
\subsection{Another one parameter scenario: curvature-induced and monotonicity}
\label{sec:curvIndMon}

\begin{figure}[h!]
    \centering    \includegraphics[width=0.7 \textwidth]{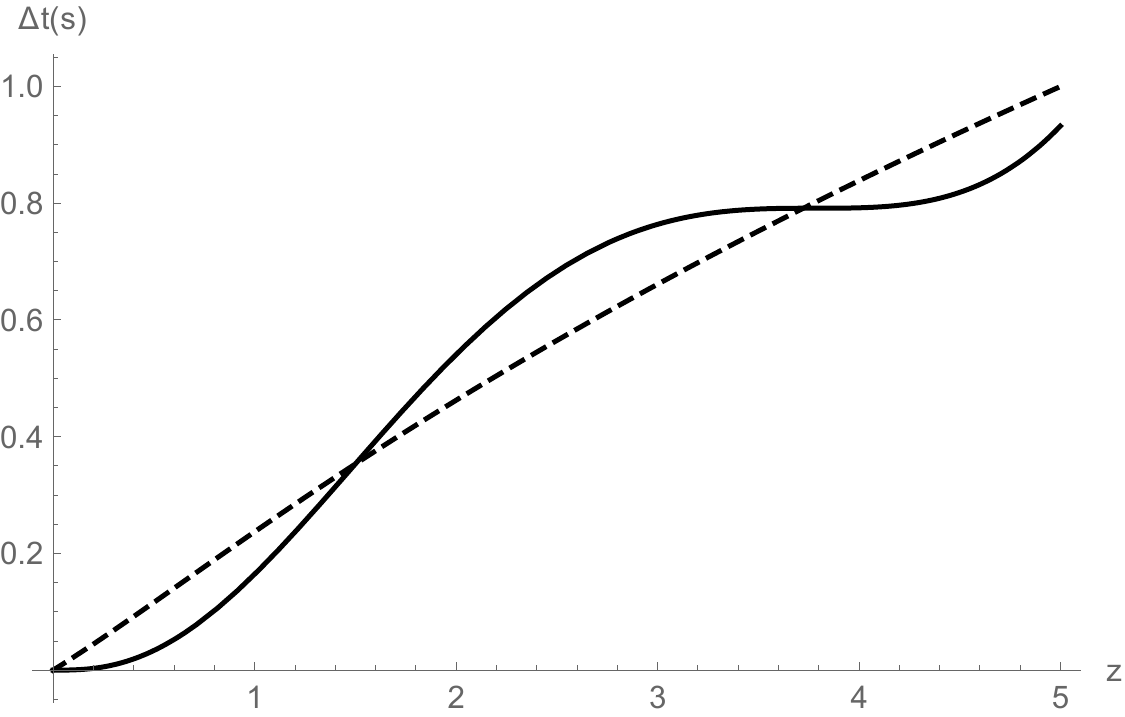}
    \caption{The continuous line corresponds to the ``curvature-induced" monotonic scenario described by~(\ref{monotoniccurind}) with $\eta_3=-1$ .
The dashed line represents the expected time delay for the Jacob-Piran case~(\ref{JacobPiran}), corresponding to setting $\eta_2=\eta_3=0$ in~(\ref{frwdelayA}); the remaining free parameter in~(\ref{frwdelayA}),
$\eta_1$, is fixed by asking that the two lines cross at $z=1.5$. As for Figure~\ref{figura1}, we set $\Delta E = 10~\text{GeV}$ for both lines. 
}
    \label{moncurvind}
\end{figure}

In the previous subsections we described some noteworthy curvature-induced scenarios where the redshift dependence of the time delay is not monotonic.
We do not see any robust argument against the lack of monotonicity, the lack of monotonicity produces no ``pathology".
Still, one might simply wonder whether monotonicity is at all possible in the curvature-induce scenario. For this purpose we must check if it is possible for the derivative of the time delay with respect to the redshift parameter $z$ to never change sign. 
We find that this condition is satisfied only when $\eta_2=-2\eta_3$. Therefore, the two conditions of monotonicity and curvature-induced effects lead to another one-parameter model. With this constraint, Eq.~(\ref{frwdelayA}) reads
\begin{equation}\label{monotoniccurind}
\begin{split}
\Delta t =\eta_3 \frac{\Delta E}{M_{Pl}} \int_{0}^{z}\frac{d\overline{z}\left(1+\overline{z}\right)}{H\left(\overline{z}\right)} \Bigg[ & -2\left(1 \!-\! \left(1 \!-\! \frac{H\left(\overline{z}\right)}{1+\bar{z}}\int_{0}^{\overline{z}}\frac{d\overline{z}'}{H\left(\overline{z}'\right)}\right)^{2}\right) \\& +\left(1 \!-\! \left(1 \!-\! \frac{H\left(\overline{z}\right)}{1+\bar{z}}\int_{0}^{\overline{z}}\frac{d\overline{z}'}{H\left(\overline{z}'\right)}\right)^{4}\right) \Bigg] .
\end{split}
\end{equation}
We illustrate the behaviour of this time delay in Figure \ref{moncurvind}. Interestingly, in this scenario there is a range of redshifts, between $\sim 3$ and $\sim 4.5$, where the time delay is approximately constant (in particular it has a stationary point at $z\sim 3.8$). This range does not depend on the value of the model parameter.

\newpage

\subsection{Monotonicity for $\eta_1 \neq 0$}
\begin{figure}[h!]
    \centering
    \includegraphics[width=0.6 \textwidth]{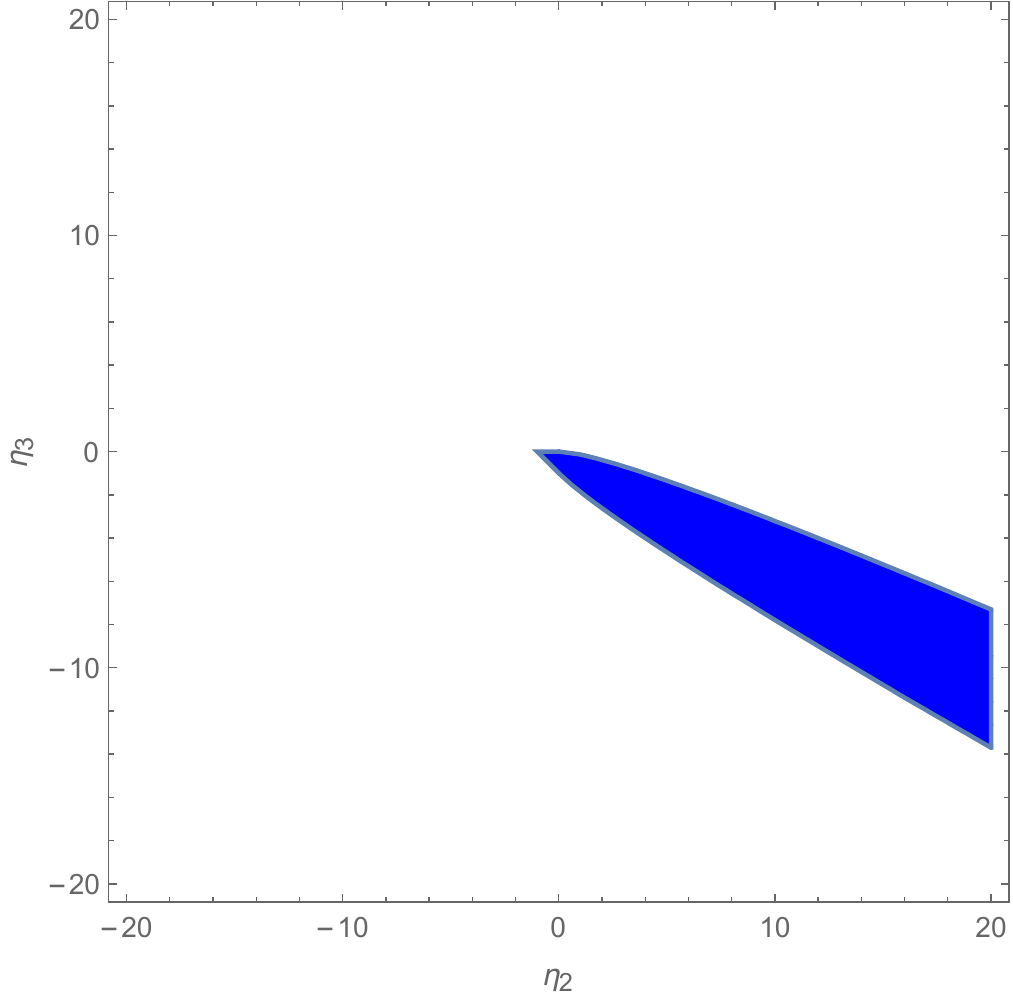}
    \caption{The blue region identifies the range of parameters $\eta_2,\eta_3$ such that the time delay of Eq.~\eqref{frwdelayA} depends monotonically on redshift when $\eta_1=1$.}
    \label{moncurvindplotnci}
\end{figure}

Having found an interesting (at least unique) scenario by requesting monotonicity in the curvature-induced case ($\eta_1 = 0$), we find appropriate to also
explore monotonicity for the most general case ($\eta_1\neq 0$).
When $\eta_1\neq 0$
the requirement of monotonicity of the time delay can be expressed by identifying a region of the parameter space \{$\eta_2/\eta_1$,
$\eta_3/\eta_1$\} where monotonicity holds. We illustrate  this  in Figure~\ref{moncurvindplotnci} by fixing $\eta_1=1$ and considering the range $\eta_2,\eta_3 \in [-20,20] $. The blue area identifies the values of $\eta_2,\eta_3$ such that $\frac{d\Delta t}{dz}\geq 0 $ for every value of the redshift.

\section{Conclusions}\label{conclusions}

We have derived the most general formula that describes the leading order time delays (assuming linear dependence on the particle energy) for ultra-relativistic particles propagating in an FLRW expanding spacetime with deformed (DSR) relativistic symmetries.
This is reported in Eq.~(\ref{frwdelayA}), which represents our main result.
We found that the requirement of relativistic consistency of the DSR scenario allows for only three possible independent forms of redshift dependence.
This is completely different from LIV scenarios, where relativistic symmetries are broken and the lack of relativistic constraints allows in principle any possible form of redshift dependence in the time-delay formula.

Considering the smallness of the Planck length and the rather poor quality of presently-obtainable data, even the exploration of the small three-parameter space of Eq.~(\ref{frwdelayA}) is a big challenge for phenomenological studies, and initially it might be necessary to focus on some specific choices of our three parameters.
We highlighted in Sec.~\ref{sec:specialCases} some choices of the three parameters which can be motivated by theoretical arguments based on the possible requirement that
the quantum gravity effects are ``curvature induced", so that the time-delay vanishes when the spacetime curvature/expansion is negligible, and the possible requirement that the total energy of a multi-particle system should be obtained with a standard linear law of addition of particle energies.
In particular, we found that combining these two possible requirements one specifies completely the redshift dependence of the effects  (Subsec.~\ref{sec:oneparam}).
Similarly, combining the curvature-induced requirement with 
a requirement of 
monotonicity of the redshift dependence also specifies
completely the redshift dependence of the effects (Subsec.~\ref{sec:curvIndMon}).
We believe that valuable first targets for phenomenology would be to discriminate 
between these two particular forms of DSR-allowed redshift dependence  
and the redshift dependence of the Jacob-Piran scenario.

\appendix

\section{de Sitter slicing for the LIV-FLRW Jacob-Piran  scenario}\label{appendix}
The derivation of FLRW travel times using the technique of ``de Sitter slicing"
was discussed in detail, mostly for DSR-relativistic scenarios, in Refs.~\cite{DSRFRW,Marciano:2010gq}.
We here focus on showing that applications of the de-Sitter slicing to the LIV Jacob-Piran scenario produce results that are equivalent to those found by redshifting the relevant momenta with the standard general relativistic scale factor. Indeed our objective is to show that also Jacob and Piran could have derived their form of redshift dependence using de Sitter slicing.

We work with comoving-time  coordinates and we 
start by considering the LIV-modified 
relationship between energy and momentum 
for a massless particle in (2D) de Sitter spacetime 
\begin{equation}
E^{2} = e^{-2Ht}p^{2} - \lambda e^{-2Ht}Ep^{2}\ ,\label{dispLIVrep}
\end{equation}
where $\lambda$ is the LIV scale, and this formula reproduces the
Jacob-Piran redshift dependence~\cite{JacobPiran} in the de Sitter
limit for which the FLRW scale factor 
is $a\left(t\right)\rightarrow e^{Ht}$,
with constant $H$. 

The speed of a massless particle can be easily obtained from (\ref{dispLIVrep})
to be (working again at first order in $\lambda$)
\begin{equation}
v\left(t\right)=\frac{\partial E}{\partial p}\simeq e^{-Ht}\left(1-\lambda e^{-Ht}p\right)\ .\label{speeddS}
\end{equation}

We want to determine the difference in  arrival times between a hard photon (a high-energy photon,
tangibly affected by LIV) and a soft photon (a low energy photon, for which the
LIV effects can be neglected), emitted simultaneously at a distant
source, traveling through a LIV-modified FLRW spacetime corresponding
to (\ref{dispLIVrep}). We consider an observer Alice local to the
event of emission (Alice's frame origin coincides with the emitter),
and an observer Bob local at the detector (Bob's frame origin coincides
with the detector), and we assume that the soft photon has been emitted
at the (comoving) Bob time $-T$. To reconstruct the trajectories
of the photons we divide the time interval $T$ between the event of emission
and the event of detection in $N$ time intervals of equal temporal
size $T/N$, such that each ``spacetime slice" is described, with good
approximation, by a constant expansion rate $H_{n}=H\left(t_{n}\right)$,
where $t_{n}$ is the initial time of the $n$-th slice, and $n=1,\dots,N$.
We consider now a set of intermediate observers Bob$_{n}$ such that
the soft photon crosses the origin of their reference frame at the
time $t_{n}$, so that $\text{Bob}_{N}=\text{Bob}$ (and Bob$_{0}$=Alice).
Each observer Bob$_{n}$, in the corresponding $n$-th slice, which
goes from $t_{n-1}$ to $t_{n}$, will describe the motion of particles
in terms of a constant expansion rate $H_{n}$, and will describe
the photons to travel with a speed $v_{n}^{B_{n}}$.

To obtain the time delay at the detector, we are interested in the
trajectory that $\text{Bob}_{N}$ assigns to the hard photon in the
final $N$-th slice (the soft photon arrives by assumption in Bob$_{N}$'s
spatial and temporal origin), which is given by
\begin{equation}
x^{B_{N}}\left(t^{B_{N}}\right)_{N}=x_{O_{A}}^{B_{N}}+\sum_{n=1}^{N-1}\int_{t_{O_{n-1}}^{B_{N}}}^{t_{O_{n}}^{B_{N}}}v_{n}^{B_{N}}dt^{B_{N}}+\int_{t_{O_{N-1}}^{B_{N}}}^{t^{B_{N}}}v_{N}^{B_{N}}dt^{B_{N}}\ ,\label{trajectoryIntegral}
\end{equation}
with $v_{n}^{B_{N}}$ the velocity that Bob$_{N}$ assigns to the
photon in the $n$th slice, and $(x_{O_{n}}^{B_{N}},t_{O_{n}}^{B_{N}})$
the coordinates that Bob$_{N}$ assigns to the photon when it crosses
the (time) origin of observer Bob$_{n}$'s frame.

To compute these quantities, it is necessary to establish the
relations between the observers' coordinates. We describe each Bob$_{n}$
as the observer connected to Alice by a set of $n$ spatial translations
followed by a set of $n$ time translations, with each $k$-th translation
characterized by the relative constant expansion rate $H_{k}$ and
finite translation parameters $\zeta_{k},\xi_{k}$, i.e. 
\begin{equation}
\left(t,x\right)^{B_{n}}=e^{-\sum_{k=1}^{n}\xi_{k}p}\triangleright e^{-\sum_{k=1}^{n}\zeta_{k}E_{H_{k}}}\triangleright\left(t,x\right)^{A}\ .\label{FiniteTransSlicing}
\end{equation}
Since in the LIV case the relativistic transformations are not deformed, one
easily finds the following relation between
Bob$_{n}$'s and Alice's coordinates 
\begin{equation}
\begin{gathered}t^{B_{n}}\left(t^{A},x^{A}\right)=t^{A}-\sum_{k=1}^{n}\zeta_{k},\\
x^{B_{n}}\left(t^{A},x^{A}\right)=e^{\sum_{k=1}^{n}H_{k}\zeta_{k}}\left(x^{A}-\sum_{k=1}^{n}\xi_{k}\right).
\end{gathered}
\label{BobnAl}
\end{equation}
The requirement for each observer Bob$_{n}$ to be along the soft
photon trajectories at the time $t_{n}$, is then ensured by imposing
that the translation parameters satisfy the conditions
\begin{equation}
\zeta_{n}=\zeta=T/N,\qquad\xi_{n}=e^{-\sum_{k=1}^{n}H_{k}\zeta_{n}}\frac{e^{H_{n}\zeta_{n}}-1}{H_{n}},\label{param}
\end{equation}
and Alice describes each $n$-th slice to be of temporal size $\zeta$
and spatial size $\xi_{n}$.

The computation of $v_{n}^{B_{N}}$ requires the use of these formulas
and a suitable matching of the scale factors $a_{n}\left(t\right)=\exp\left(H_{n}t\right)$
at the junction of each slice, after which one obtains
\begin{equation}
v_{n}^{B_{N}}=\frac{1}{a_{n}^{B_{N}}\left(t^{B_{N}}\right)}\left(1-\lambda\frac{p^{B_{N}}}{a_{n}^{B_{N}}\left(t^{B_{N}}\right)}\right)\ ,\label{vnBN}
\end{equation}
where
\begin{equation}
a_{n}^{B_{N}}\left(t^{B_{N}}\right)=e^{-\sum_{k=n+1}^{N}H_{k}\zeta}e^{\left(N-n\right)H_{n}\zeta}e^{H_{n}t^{B_{N}}}\ .
\end{equation}
The velocity (\ref{vnBN}) can be easily integrated in each slice
in (\ref{trajectoryIntegral}), where we take
\begin{equation}
t_{O_{n}}^{B_{N}}=t^{B_{N}}\left(t^{B_{n}}=0\right)=-\left(N-n\right)\zeta\ ,
\end{equation}
and considering that, combining (\ref{BobnAl}) and (\ref{param}),
one has that 
\begin{equation}
x_{O_{A}}^{B_{N}}=x^{B_{N}}\left(x^{A}=0,t^{A}=0\right)=-\sum_{k=1}^{N}e^{\sum_{s=k}^{N}H_{s}\zeta}\frac{1-e^{-H_{k}\zeta}}{H_{k}}\ ,
\end{equation}
the trajectory in the $N$-th slice is given by
\begin{equation}
x^{B_{N}}\left(t^{B_{N}}\right)=\frac{1-e^{-H_{N}t^{B_{N}}}}{H_{N}}-\lambda p^{B_{N}}\left(\sum_{n=1}^{N}e^{2\sum_{k=n}^{N}H_{k}\zeta}\frac{1-e^{-2H_{n}\zeta}}{2H_{n}}+\frac{1-e^{-2H_{N}t^{B_{N}}}}{2H_{N}}\right)\ .
\end{equation}
From the trajectory we obtain the hard-photon time delay (at first
order in $\lambda$) by solving for $t^{B_{N}}\left(x^{B_{N}}=0\right)$,
\begin{equation}
\Delta t^{B_{N}}=\lambda p^{B_{N}}\sum_{n=1}^{N}e^{2\sum_{k=n}^{N}H_{k}\zeta}\frac{1-e^{-2H_{n}\zeta}}{2H_{n}}\ .
\end{equation}
We take now the limit $N\rightarrow\infty$, in which the slices
are infinitesimally small. Using the
formulas ($\zeta=T/N$)
\begin{equation}
\sum_{k=n_{i}+1}^{n_{f}}\zeta\rightarrow\int_{t_{n_{i}}}^{t_{n_{f}}}dt\ ,
\end{equation}
and, noticing that, since $H(t)\!=\!\dot{a}(t)/a(t)$, $a\left(t_{f}\right)/a\left(t_{i}\right)=\exp\left(\int_{t_{i}}^{t_{f}}dt\ H\left(t\right)\right)$, and
\begin{equation}
e^{\sum_{s=k+1}^{n}H_{s}\zeta}\rightarrow\frac{a\left(t_{n}\right)}{a\left(t_{k}\right)}\ ,
\end{equation}
we obtain
\begin{equation}
\Delta t\rightarrow\lambda p_{h}\int_{-T}^{0}\frac{dt}{a^{2}\left(t\right)}\ ,
\end{equation}
where we denoted by $p_{h}$ the momentum of the hard particle observed
at the detector, we considered that for $N\!\rightarrow\!\infty$ one
has that $\frac{e^{2H_{k}\zeta}-1}{2H_{k}}\!\rightarrow\zeta$, that
$t_{0}=-T$, $t_{N}=0$, and that $a\left(t_{N}\right)=a\left(0\right)=1$.
Finally, we can rewrite the delay in terms of the redshift of the
source $z\equiv z(-T)$, noticing that, for $\bar{z}\equiv z(t)$, $a\left(t\right)=1/(1+\bar{z})$
and $dt=-d\bar{z}/(H(\bar{z})(1+\bar{z}))$, so that
\begin{equation}
\Delta t=\lambda p_h\int_{0}^{z}\frac{d\bar{z}\left(1+\bar{z}\right)}{H\left(\bar{z}\right)}\ ,
\end{equation}
which indeed coincides with the formula obtained by Jacob and Piran in~\cite{JacobPiran}.

\acknowledgments

G.A.-C., D.F. and G.G. acknowledge financial support by the Programme STAR Plus, funded by Federico II University and Compagnia di San Paolo, and by the MIUR, PRIN 2017 grant 20179ZF5KS. G.R.’s work on this project was supported by the National Science Centre grant
2019/33/B/ST2/00050. G.G and G.R.  thank Perimeter Institute for their hospitality in May 2023, where the final stages of the project were completed. Research at the Perimeter Institute for Theoretical Physics is supported in part by the Government of Canada through NSERC and by the Province of Ontario through MRI.  This work contributes to the European Union COST Action CA18108 {\it Quantum gravity phenomenology in the multi-messenger approach.}



\begin{thebibliography}{99}

\bibitem{GACreview}
G.~Amelino-Camelia,
``Quantum-Spacetime Phenomenology,''
Living Rev. Rel. \textbf{16} (2013), 5
[arXiv:0806.0339 [gr-qc]].

\bibitem{MattinglyReview}
D.~Mattingly,
``Modern tests of Lorentz invariance,''
Living Rev. Rel. \textbf{8} (2005), 5
[arXiv:gr-qc/0502097 [gr-qc]].

\bibitem{COSTReview}
A.~Addazi, J.~Alvarez-Muniz, R.~Alves Batista, G.~Amelino-Camelia, V.~Antonelli, M.~Arzano, M.~Asorey, J.~L.~Atteia, S.~Bahamonde and F.~Bajardi, \textit{et al.}
``Quantum gravity phenomenology at the dawn of the multi-messenger era\textemdash{}A review,''
Prog. Part. Nucl. Phys. \textbf{125} (2022), 103948
[arXiv:2111.05659 [hep-ph]].

\bibitem{GRBnature}
G.~Amelino-Camelia, J.~R.~Ellis, N.~E.~Mavromatos, D.~V.~Nanopoulos and S.~Sarkar,
``Tests of quantum gravity from observations of gamma-ray bursts,''
Nature \textbf{393} (1998), 763-765
[arXiv:astro-ph/9712103 [astro-ph]].

\bibitem{Alfaro:1999wd}
J.~Alfaro, H.~A.~Morales-Tecotl and L.~F.~Urrutia,
Phys. Rev. Lett. \textbf{84} (2000), 2318-2321
[arXiv:gr-qc/9909079 [gr-qc]].
\bibitem{Gambini:1998it}
R.~Gambini and J.~Pullin,
Phys. Rev. D \textbf{59} (1999), 124021
[arXiv:gr-qc/9809038 [gr-qc]].
\bibitem{Amelino-Camelia:2000stu}
G.~Amelino-Camelia,
``Relativity in space-times with short distance structure governed by an observer independent (Planckian) length scale,''
Int. J. Mod. Phys. D \textbf{11} (2002), 35-60
[arXiv:gr-qc/0012051 [gr-qc]].
\bibitem{Magueijo:2002am}
J.~Magueijo and L.~Smolin,
Phys. Rev. D \textbf{67} (2003), 044017
[arXiv:gr-qc/0207085 [gr-qc]].
\bibitem{Kowalski-Glikman:2002iba}
J.~Kowalski-Glikman and S.~Nowak,
Phys. Lett. B \textbf{539} (2002), 126-132
[arXiv:hep-th/0203040 [hep-th]].

\bibitem{PiranMartinez}
M.~Rodriguez Martinez and T.~Piran,
``Constraining Lorentz violations with gamma-ray bursts,''
JCAP \textbf{04} (2006), 006
[arXiv:astro-ph/0601219 [astro-ph]].

\bibitem{Pfeifer:2018pty}
C.~Pfeifer,
``Redshift and lateshift from homogeneous and isotropic modified dispersion relations,''
Phys. Lett. B \textbf{780} (2018), 246-250
[arXiv:1802.00058 [gr-qc]].
\bibitem{Lobo:2016xzq}
I.~P.~Lobo, N.~Loret and F.~Nettel,
``Investigation of Finsler geometry as a generalization to curved spacetime of Planck-scale-deformed relativity in the de Sitter case,''
Phys. Rev. D \textbf{95} (2017) no.4, 046015
[arXiv:1611.04995 [gr-qc]].



\bibitem{JacobPiran}
U.~Jacob and T.~Piran,
``Lorentz-violation-induced arrival delays of cosmological particles,''
JCAP \textbf{01} (2008), 031
[arXiv:0712.2170 [astro-ph]].
\bibitem{EllisMavroFRWdelay}
J.~R.~Ellis, N.~E.~Mavromatos, D.~V.~Nanopoulos and A.~S.~Sakharov,
``Quantum-gravity analysis of gamma-ray bursts using wavelets,''
Astron. Astrophys. \textbf{402} (2003), 409-424
[arXiv:astro-ph/0210124 [astro-ph]].
\bibitem{Planck:2018vyg}
Aghanim N. \textit{et al.}
Planck 2018 results. VI. Cosmological parameters.
{\it Astron. Astrophys.} \textbf{641}, A6 (2020),
[erratum: {\it Astron. Astrophys.} \textbf{652}, C4 (2021)].
[arXiv:1807.06209 [astro-ph.CO]].

\bibitem{DSRFRW}
G.~Rosati, G.~Amelino-Camelia, A.~Marciano and M.~Matassa,
``Planck-scale-modified dispersion relations in FRW spacetime,''
Phys. Rev. D \textbf{92} (2015) no.12, 124042
[arXiv:1507.02056 [hep-th]].

\bibitem{Bolmont:2022yad}
J.~Bolmont, S.~Caroff, M.~Gaug, A.~Gent, A.~Jacholkowska, D.~Kerszberg, C.~Levy, T.~Lin, M.~Martinez and L.~Nogu\'es, \textit{et al.}
``First Combined Study on Lorentz Invariance Violation from Observations of Energy-dependent Time Delays from Multiple-type Gamma-Ray Sources. I. Motivation, Method Description, and Validation through Simulations of H.E.S.S., MAGIC, and VERITAS Data Sets,''
Astrophys. J. \textbf{930} (2022) no.1, 75
[arXiv:2201.02087 [astro-ph.HE]].
\bibitem{DSRdS}
G.~Amelino-Camelia, A.~Marciano, M.~Matassa and G.~Rosati,
``Deformed Lorentz symmetry and relative locality in a curved/expanding spacetime,''
Phys. Rev. D \textbf{86} (2012), 124035
[arXiv:1206.5315 [hep-th]].

\bibitem{Marciano:2010gq}
A.~Marciano, G.~Amelino-Camelia, N.~R.~Bruno, G.~Gubitosi, G.~Mandanici and A.~Melchiorri,
``Interplay between curvature and Planck-scale effects in astrophysics and cosmology,''
JCAP \textbf{06} (2010), 030
[arXiv:1004.1110 [gr-qc]].

\bibitem{Barcaroli:2015eqe}
L.~Barcaroli and G.~Gubitosi,
``Kinematics of particles with quantum-de Sitter-inspired symmetries,''
Phys. Rev. D \textbf{93} (2016) no.12, 124063
[arXiv:1512.03462 [gr-qc]].

\bibitem{GACSmolinqdeSitter}
G.~Amelino-Camelia, L.~Smolin and A.~Starodubtsev,
``Quantum symmetry, the cosmological constant and Planck scale phenomenology,''
Class. Quant. Grav. \textbf{21} (2004), 3095-3110
[arXiv:hep-th/0306134 [hep-th]].


\bibitem{taming}
G.~Amelino-Camelia, M.~Matassa, F.~Mercati and G.~Rosati,
``Taming Nonlocality in Theories with Planck-Scale Deformed Lorentz Symmetry,''
Phys. Rev. Lett. \textbf{106} (2011), 071301
[arXiv:1006.2126 [gr-qc]].

\bibitem{kbob}
G.~Amelino-Camelia, N.~Loret and G.~Rosati,
``Speed of particles and a relativity of locality in $\kappa$-Minkowski quantum spacetime,''
Phys. Lett. B \textbf{700} (2011), 150-156
[arXiv:1102.4637 [hep-th]].

\bibitem{Mignemi:2019yzn}
S.~Mignemi and G.~Rosati,
``Physical velocity of particles in relativistic curved momentum space,''
Mod. Phys. Lett. A \textbf{35} (2020) no.22, 2050180
[arXiv:1909.09173 [gr-qc]].

\bibitem{principle}
G.~Amelino-Camelia, L.~Freidel, J.~Kowalski-Glikman and L.~Smolin,
``The principle of relative locality,''
Phys. Rev. D \textbf{84} (2011), 084010
[arXiv:1101.0931 [hep-th]].

\bibitem{Amelino-Camelia:2013uya}
G.~Amelino-Camelia, L.~Barcaroli, G.~Gubitosi and N.~Loret,
``Dual redshift on Planck-scale-curved momentum spaces,''
Class. Quant. Grav. \textbf{30} (2013), 235002
[arXiv:1305.5062 [gr-qc]].



\bibitem{aschieri09}
P.~Aschieri, A.~Borowiec and A.~Pacho{\l},
``Dispersion relations in $\kappa$-noncommutative cosmology,''
JCAP \textbf{04} (2021), 025
[arXiv:2009.01051 [gr-qc]].

\bibitem{bianchiRovelli}
E.~Bianchi and C.~Rovelli,
``A Note on the geometrical interpretation of quantum groups and non-commutative spaces in gravity,''
Phys. Rev. D \textbf{84} (2011), 027502.

\bibitem{curvInducedLIV}
G.~Amelino-Camelia, G.~Rosati and S.~Bedi\'c,
``Phenomenology of curvature-induced quantum-gravity effects,''
Phys. Lett. B \textbf{820} (2021), 136595
[arXiv:2012.07790 [gr-qc]].



\bibitem{Amelino-Camelia:2011gae}
G.~Amelino-Camelia,
``On the fate of Lorentz symmetry in relative-locality momentum spaces,''
Phys. Rev. D \textbf{85} (2012), 084034
[arXiv:1110.5081 [hep-th]].

\bibitem{Amelino-Camelia:2013sba}
G.~Amelino-Camelia, G.~Gubitosi and G.~Palmisano,
``Pathways to relativistic curved momentum spaces: de Sitter case study,''
Int. J. Mod. Phys. D \textbf{25} (2016) no.02, 1650027
[arXiv:1307.7988 [gr-qc]].
\bibitem{Amelino-Camelia:2023rkg}
G.~Amelino-Camelia, G.~Fabiano and D.~Frattulillo,
``Total momentum and other Noether charges for particles interacting in a quantum spacetime,''
[arXiv:2302.08569 [hep-th]].



\bibitem{Ball19943D}
  A.~Ballesteros, F.~J.~Herranz, M.~A.~del Olmo and M.~Santander,
  ``Quantum (2 + 1) kinematical algebras: a global approach''
  J.\ Math.\ Phys.\  {\bf 27} (1994) 1283

\bibitem{Ballesteros:2014kaa}
A.~Ballesteros, F.~J.~Herranz, C.~Meusburger and P.~Naranjo,
``Twisted (2+1) $\kappa$-AdS Algebra, Drinfel'd Doubles and Non-Commutative Spacetimes,''
SIGMA \textbf{10} (2014), 052
[arXiv:1403.4773 [math-ph]].

\bibitem{jack3Dgravity}
G.~Rosati,
``$\kappa$\textendash{}de Sitter and $\kappa$-Poincar\'e symmetries emerging from Chern-Simons (2+1)D gravity with a cosmological constant,''
Phys. Rev. D \textbf{96} (2017) no.6, 066027
[arXiv:1706.02868 [hep-th]].

\bibitem{Ballesteros:2017pdw}
A.~Ballesteros, G.~Gubitosi, I.~Guti\'errez-Sagredo and F.~J.~Herranz,
``Curved momentum spaces from quantum (anti\textendash{})de Sitter groups in ( 3+1 ) dimensions,''
Phys. Rev. D \textbf{97} (2018) no.10, 106024
[arXiv:1711.05050 [hep-th]].

\end{thebibliography}


\end{document}